\documentclass[english,pre,showpacs,floatfix,manuscript]{revtex4}
\usepackage{babel}
\usepackage{epsfig}
\usepackage{amsmath}
\usepackage{amssymb}

\begin{document}

\title{Directed transient long-range transport in a slowly driven Hamiltonian system of interacting particles}

\author{D. Hennig}

\affiliation{Institut f\"{u}r Physik, Humboldt-Universit\"{a}t zu Berlin, Newtonstr. 15, 12489 Berlin, Germany}

\begin{abstract}
\noindent We study the Hamiltonian dynamics of a one-dimensional
chain of linearly coupled particles in a spatially
periodic potential which is subjected to a time-periodic mono-frequency external field. The average over time and space of the
related force vanishes and hence, the
system is effectively without bias which excludes any ratchet effect. We pay special attention to
the escape of the entire chain when initially all of its units are
distributed in a potential well. Moreover for an escaping chain we explore the
possibility of the successive generation of a directed flow based on 
large accelerations. We find
that for adiabatic slope-modulations due to the ac-field transient long-range transport dynamics arises whose direction 
is governed by the initial phase of the modulation. 
Most strikingly, that for the driven
many particle Hamiltonian system
directed collective motion 
is observed provides evidence for the existence of families of transporting invariant tori confining orbits in ballistic channels in the high 
dimensional phase spaces.
\end{abstract}


\maketitle

\section{Introduction}

Transport phenomena play a fundamental role in many physical systems. In this context the so called washboard potential 
due to its ubiquity and simplicity establishes the prototype of a periodic potential being employed in a number of applications 
including Josephson junctions \cite{Josephson}, charge density waves \cite{charge}, superionic conductors \cite{superionic}, rotation of dipoles in external fields \cite{dipoles}, phase-locked loops \cite{loops} and diffusion of dimers on surfaces \cite{surfaces,Braun,Pijper,Braun1,Heisalu,Patriarca,Fusco} to quote a few. Applying an external time periodic driving to the washboard potential, 
interesting effects such as phase-locking, hysteresis \cite{hysteresis} and stochastic resonance \cite{stochastic} are found. Recent investigations have dealt with the Hamiltonian dynamics of individual particles evolving in a washboard  
potential whose slope is time-periodically varied by
 a weak external monochromatic modulation field \cite{Yevtushenko,Soskin,single}. Remarkably, for the corresponding one-and-a-half degree of freedom Hamiltonian system it has been
demonstrated that adiabatic modulations of the slope of the potential lead to the generation of transient transport dynamics related with enormous directed particle flow. Explanation for this behavior has been given in terms of the underlying phase space structure
of the externally driven one degree of freedom system 
promoting the motion in ballistic channels \cite{ballistic}. 
Here we consider an extension of previous work on individual particle movement and study the case when the (many) particles interact due to linear couplings constituting a one-dimensional chain. 
Motion takes place in a phase space for which details of its intricate structures remain elusive not at least due to the high dimensionality.
Whether in systems with a large number of (microscopic) degrees of freedom such (macroscopic) behavior as collective motion leading to a directed flow emanates from high dimensional dynamics is not obvious. In systems with at least three degrees of freedom  
irregular dynamics in the form of Arnold diffusion arises
\cite{Arnolddiff}-\cite{Lieberman}. This is possible
because the KAM tori do not separate the phase space so that the various chaotic regions always overlap and in principle orbits can wander along the entire stochastic layer permeating the whole phase space. 
Let us recall that in one-and-a-half degree of freedom 
weakly nonintegrable Hamiltonian systems the motion can stick to the hierarchical structure of persisting tori and islands inside stochastic layers being able to constitute the above mentioned ballistic channels which support directed transport. The question then is whether such trapping regions persist in many
degrees of freedom Hamiltonian systems supporting also directed transport. 
It is the objective of the current work  to demonstrate that even in a system of many coupled particles the generation
of a directed flow going along with collective motion is possible indeed. 

\vspace*{0.5cm}

\section{The Driven System Of Interacting Particles}

We study a one-dimensional chain system consisting of
linearly coupled particles with Hamiltonian of the following form
\begin{eqnarray}
H&=&\sum_{n=1}^{N}\left[\frac{p_n^{2}}{2}+U_0(q_n)+U_1(q_n,t)\right]\nonumber\\
&+&\frac{\kappa}{2}\sum_{n=1}^{N-1}\left(q_{n+1}-q_{n}\right)^2
\,,\label{equation:Hamiltonian}
\end{eqnarray}
wherein $p_n$ and $q_n$ denote the canonically conjugate momenta and
positions of the particles evolving in the periodic, spatially-symmetric (washboard) 
potential of unit period, i.e.,
\begin{equation}
U_0(q)=U_0(q+1)=-\cos(2\pi q)/(2\pi)\,.
\end{equation}
The external, time-dependent forcing field
\begin{equation}
U_1(q,t)=-F\sin({\Omega \,t+\Theta_0})q
\end{equation}
causes time-periodic modulations of the slope of the potential. It
has to be stressed that there is no bias force involved in the sense that
the following average over time and space vanishes, i.e.
\begin{equation}
\int\limits_{0}^{1}dq\int\limits_0^{T=2\pi/\Omega}dt
\frac{\partial U(q,t)}{\partial q}=0\,,
\end{equation}
with $U(q,t)=U_0(q)+U_1(q,t)$. The particles interact linearly
with coupling strength $\kappa$.
Remarkably, as pointed out in  prior literature \cite{Yevtushenko}
in the limiting case of uncoupled particles, i.e. $\kappa=0$, there results an (unexpected)
asymmetry of the flux of particles, emanating from one potential
well, and flowing to the left and right potential wells which
indicates the existence of directed transport without breaking the
reflection symmetry in space and time in this system. One reason for
the occurrence of phase-dependent directed transport is the lowering
of the symmetry of the flow in phase space by the ac-field where
this asymmetry vanishes only for specific values of the initial
phase $\Theta_0$ \cite{Yevtushenko}. In Ref.~\cite{Soskin}
the authors report further on this exceptional situation and show
that directed transport is sustained on fairly long time scales
despite the presence of chaos. In particular it has been
demonstrated that for sufficiently small forcing frequencies,
$\Omega \ll 1$, the width of the arising chaotic layer diverges
leading to a strong enhancement of the chaotic transport
\cite{Soskin}.

The equations of motion derived from the Hamiltonian in Eqs. (\ref{equation:Hamiltonian}) read as
\begin{equation}
\frac{d^2q_n}{dt^2}\,+\,\sin(2\pi q_n)\,=\,F\sin(\Omega\, t +\Theta_0)+\kappa[q_{n+1}+q_{n-1}-2q_n]\,\,.\label{eq:qdot}
\end{equation}
The analysis in the current paper deals with zero initial phase of the external force term, viz. $\Theta_0=0$.
The influence of the phase $\Theta_0$ on the generation of directed motion in the periodic potential $U_0(q)$ has been studied in  \cite{single}. 
We stress that averaging over the initial phase $\Theta_0$ yields vanishing net current in the periodically driven system.

\vspace*{0.5cm}

\section{The Case $N=2$}

Before we embark on the study of systems
comprising of many degrees of freedom let us first consider the case
of two interacting particles. For a discussion of the dynamics it is convenient to introduce the
following canonical change of variables induced by the generating
function: $S=\frac{1}{2}(q_1+q_2)P_x+\frac{1}{2}(q_1-q_2)P_y$
relating the old and new variables as
\begin{eqnarray}
p_1&=&\frac{1}{2}(P_x+P_y)\,,\qquad p_2=\frac{1}{2}(P_x-P_y)\,,\\
Q_x&=&\frac{1}{2}(q_1+q_2)\,,\qquad Q_y=\frac{1}{2}(q_1-q_2)\,.
\end{eqnarray}
The Hamiltonian expressed in the new variables becomes
\begin{eqnarray}
H&=&\frac{1}{4}(P_x^2+P_y^2)-\frac{1}{\pi}\cos(2\pi Q_x)\cos(2\pi
Q_y)\nonumber\\ &+&2\kappa Q_y^2-2F\sin(\Omega t)Q_x\,.
\end{eqnarray}
The corresponding equations of motion are given by
\begin{eqnarray}
\ddot{Q}_x&=&-\cos(2\pi Q_y)\sin(2\pi Q_x)\label{eq:qx}\nonumber\\
&+&F\sin(\Omega t)\,,\\ \ddot{Q}_y&=&-\cos(2\pi
Q_x)\sin(2\pi Q_y)\nonumber\\ &-&2\kappa Q_y\,.
\end{eqnarray}

Evidently, the impact of the external modulation and the linear
coupling occurs in separate equations. The coupling between the $Q_x$ and $Q_y$ degree of freedom 
results from parametric modulations of the respective potential force term.
Apart from that,
equation (\ref{eq:qx}) for the mean value variable $Q_x$ is equivalent to that
of the system of noninteracting particles. Thus, the external
modulation may lead to an escape of particles from a potential
well initiating ongoing rotational motion of the mean value $Q_x$
equivalent to the behavior observed in the study of noninteracting particles
\cite{single}. Distinct to that the amplitude of the difference variable $Q_y$
is bounded due to the harmonic restoring
force associated with the $\kappa-$term. 

\begin{figure}
\includegraphics[scale=0.45]{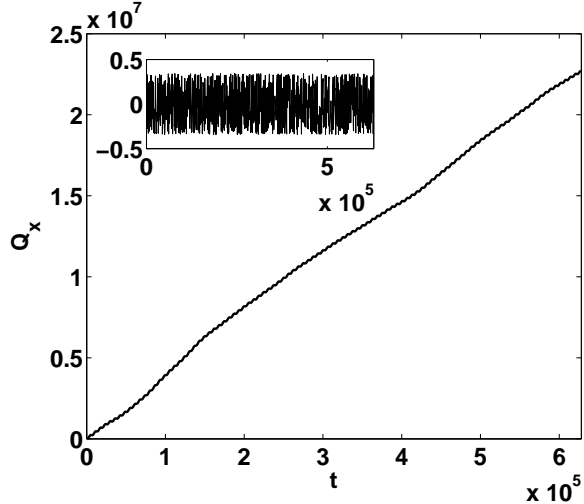}
\caption{Two units: Time evolution of the variable
$Q_x=(q_1+q_2)/2$ in the driven case of $F=0.05$. 
The remaining parameter
values are $\kappa=0.1$, $\Omega=10^{-3}$ and
$\Theta_0=0$. The initial conditions are $q_1(0)=0.337$, $q_2(0)=0.347$ and $p_1(0)=p_2(0)=0$. 
Note the large scale on the ordinate. 
The inset shows
bounded chaotic motions of $Q_x$ in the undriven case $F=0$.} \label{fig:Fig1}
\end{figure}
In our dynamical studies the initial conditions are chosen such
that the units are initially contained in a potential well.
Furthermore, in the uncoupled case, i.e $\kappa=0$, escape from
the potential well induced by the driving force, as discussed in
\cite{single}, is excluded. To ensure trapping in the driven but uncoupled dynamics the orbits have to lie fairly deep inside the separatrix where there is a large island of stability and the dynamics is still regular. 

Likewise, without the {\it weak} external ac-field and when the units are coupled the chain is supposed to remain
trapped in the potential well despite the arising chaotic motion. The latter situation is illustrated
in the inset in the top panel of Fig.~\ref{fig:Fig1}. The simulation
time interval corresponds to a
hundred times of the period duration of the external modulation
$T=2\pi/\Omega$ with $\Omega=10^{-3}$.

\begin{figure}
\includegraphics[scale=0.45]{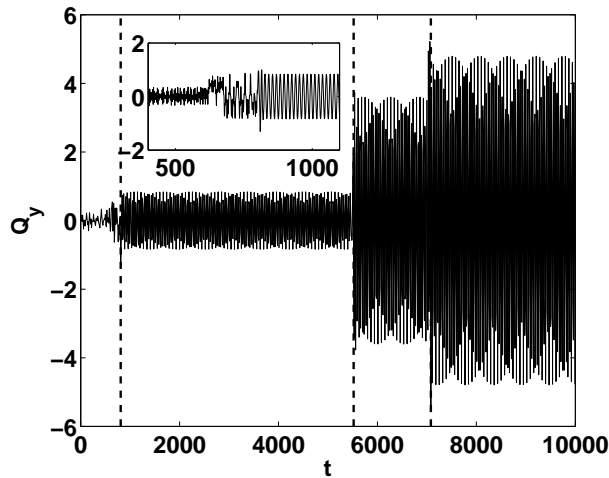}
\caption{Illustration of the changes in amplitude of the
coordinate $Q_y=(q_1-q_2)/2$ in the time interval $0\le t \le 10^4$. The initial conditions and parameter values are those of the driven case in
Fig.~\ref{fig:Fig1}. Vertical dashed lines mark the instants of
time when the amplitude (torus-torus) transitions occur. The inset displays the
transition from irregular motion prior to escape to quasi-regular
periodic motion after escape.} \label{fig:Fig2}
\end{figure}

However, when the interacting particles are exerted also to the
external ac-force escape of the particles from the
potential well occurs and results such as those
depicted in Fig.~\ref{fig:Fig1} are obtained. Interestingly, for a
slow modulation with frequency $\Omega=10^{-3}$, after escape from the potential well 
taking place at $t\simeq 800$, there occurs
long-range directed transport of the particles equivalent to the
behavior reported in \cite{single} for the system of noninteracting particles.
Notice the huge distance covered by the particles.
As reflected in the evolution of $Q_y$ (cf. inset in 
Fig.~\ref{fig:Fig2}) prior to escape  the motion is irregular whereas after escape $Q_y$ oscillates seemingly regularly indicating capture of the motion near an invariant torus (see further below). That is, the
individual coordinates $q_1$ and $q_2$ perform 
oscillations with a mutual phase lag. 
Moreover, as Fig.~\ref{fig:Fig2} reveals the $Q_y$
quasi-stable oscillations with a certain maximal amplitude are sustained over a
specific time interval at the end of which a rather abrupt
change of the amplitude occurs and the trajectory comes to lie close to another torus. On a finer scale one recognizes
that the transition from one torus to another one 
is accompanied by a chaotic interlude (see also \cite{Zumofen}).

\begin{figure}
\includegraphics[scale=0.45]{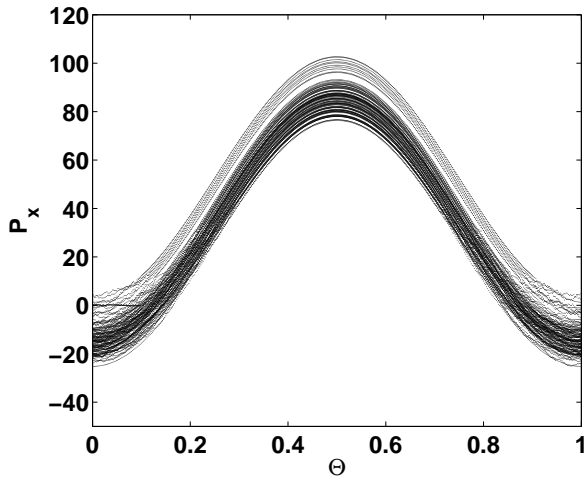}
\caption{Poincar\'{e}-plot presented in the $\Theta-P_x-$plane.
The phase variable $\Theta$ is given $\mod(2\pi)$. 
The parameter values are $\kappa=0.1$, $F=0.05$, $\Omega=10^{-3}$ and $\Theta_0=0$.}
\label{fig:Fig3}
\end{figure}

For further illustration of the dynamics in the five dimensional
phase space spanned by the variables
$(P_x,P_y,Q_x,Q_y,\Theta=\Omega t)$ we utilize
Poincar\'{e}-plots where the cross-section is determined as
follows:
\begin{equation}
\Sigma:=\left\{(P_x,Q_x,\Theta) | Q_y=0, P_y>0\right\}\,.
\end{equation}
We represent in Fig.~\ref{fig:Fig3} the Poincar\'{e}-plot in the $\Theta-P_x-$plane corresponding to the dynamics of the adiabatically driven system shown in Fig.~\ref{fig:Fig1}. The momentum $P_x$ experiences vast alterations in dependence of
the value of the phase $\Theta$. To be precise, starting from an
non-inclined potential, i.e. $\Theta_0=0$, the force term
$F(t)=F\sin(\Omega t)$ produces a negative inclination of the potential
in the interval $0< t \le \pi/\Omega$. It happens that at some
instant of time $t_{escape} \le \pi/(2\Omega)$ the coordinate
$Q_x$ overcomes the potential barrier and escapes from the well.
Hence there remains
a time interval $[t_{escape},\pi/\Omega]$ during which the
particle still experiences a force with positive sign which
accelerates further the motion in the right direction towards
increasing values of $Q_x$. For times $\pi/\Omega<t\le
2\pi/\Omega$ the force
acts in the opposite direction. In particular for
$2\pi/\Omega-t_{escape}\le t\le 2\pi/\Omega$ the momentum evolves
with its sign reversed compared to the previous acceleration
period. With this we can estimate the gain in momentum as follows
\begin{eqnarray}
\Delta P_x&=&
\left(\int_{t_{escape}}^{\pi/\Omega}+\int_{2\pi/{\Omega}-t_{escape}}^{2\pi/\Omega}\right)d\tau\dot{p}\nonumber\\&=&
\left(\int_{t_{escape}}^{\pi/\Omega}+\int_{2\pi/{\Omega}-t_{escape}}^{2\pi/\Omega}\right)
\nonumber\\ &\times& d\tau\left[-\sin(2\pi Q_x)\cos(2\pi
Q_y)+F\sin(\Omega\tau)\right]\,.
\end{eqnarray}
For small $\Omega$ the rapidly oscillating part connected with the
first term in the integral averages to zero
and we find
\begin{equation}
\Delta P_x= 2\frac{F}{\Omega}\cos(\Omega\, t_{escape})\,.\label{equation:pmax}
\end{equation}
In general, the smaller $\Omega$ the higher is the gain in momentum.
In principle, for a sufficiently small frequency $\Omega$ the gain can become arbitrarily large (see also \cite{Soskin}).
With the choice $\Theta_0=\pi$ for the initial phase of the modulation term the direction of the motion can be reversed.
In contrast in the undriven case, where the chaotic
motion remains trapped in the potential well (as depicted in the inset of Fig.~\ref{fig:Fig1}), the momentum
variable $P_x$ accordingly stays within the range determined by the extension in momentum direction of the corresponding separatrix loop (not shown). 

\begin{figure}
\includegraphics[scale=0.45]{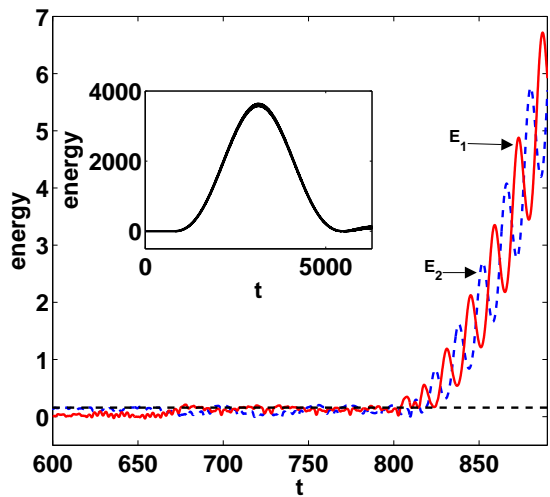}
\caption{Temporal behavior of the partial energies in the driven system of two coupled units around the trapping-detrapping transition.
Initial conditions and parameter values as in Fig.~\ref{fig:Fig1}. The dashed horizontal line marks the separatrix energy $E_s=1/(2\pi)$. The inset illustrates the drastic growth of the energies during a period $T=2\pi/\Omega$.}
\label{fig:Fig4}
\end{figure}

The temporal behavior of the partial energies $E_{1,2}=0.5
p_{1,2}^2+U_0(q_{1,2})$ is illustrated in Fig.~\ref{fig:Fig4}. For
times $t\le 800$ the energies do not considerably exceed the
separatrix level, $E_s=1/(2\pi)$, which is marked by the dashed
horizontal line. Escape from the potential well takes place at
$t_{escape}\simeq 800$ giving rise to significant increase of the
energies reaching a maximal value being equivalent to $10^5$ times the energetic barrier height $\Delta E=1/\pi$.
The two partial energies, and likewise the coordinates and
momenta, perform oscillations with mutual phase lag (see also further
above). The inset displays the temporal evolution of the energies
over one period of the external modulation term where one 
cannot discern the individual trajectories on this scale.

In terms of phase space structure we recall that in the case 
of individual particles (i.e. $\kappa=0$) in the corresponding adiabatically driven system of one-and-a-half degree of freedom  there
arises a broad stochastic layer from the connection of various zones of instability due to resonance overlap.
Non-contractible KAM tori confine the stochastic layer from below and above and form impenetrable barriers
for motion in phase space. For the weakly nonintegrable system the chaotic sea contains still
islands of
regular motion. Provided these islands possess
non-zero winding numbers orbits with initial condition inside such an island facilitate transport. Moreover, the motion
around these islands is characterized by the stickiness to them that
can lead to trapping of the trajectory for a long time \cite{Zaslavsky},\cite{MacKay}. This is due the intricate structure of
the stochastic layer where close to resonances at the boundary between
regular and chaotic regions there exists a hierarchy of smaller and
smaller islands and surrounding cantori.  The latter can severely restrict the transport 
in phase space and thus effectively partition the chaotic layer \cite{MacKay}.
It seems that the cantori are the less leaky the smaller the modulation frequency $\Omega$. Hence they  
form almost impenetrable barriers that confine trajectories for a very long but {\it transient} period. One should 
remark that eventually this transient period of directed motion terminates because the trajectory escapes through one of the holes in the cantori and
accesses other regions of the chaotic layer.  Therefore the motion does not necessarily proceed unidirectionally and
unless the trajectory gets captured by ballistic channels \cite{ballistic}
it itinerates within the chaotic layer going along with changes of the direction of motion.

For nonlinear Hamiltonian systems with $N\ge 2$ degrees of freedom 
only a few numerical results addressing the existence of an enhanced trapping regime are known \cite{Kantz}. 
It is supposed that the role played by cantori in driven systems with $N=1$ is played by families of $N-$dimensional 
tori, constituting partial barriers in the $2N-$dimensional phase space, where the chaotic trajectory can stick to \cite{Kantz}.
On the other hand Arnold diffusion is possible and hence in principle a chaotic trajectory, 
wandering along the entire stochastic layer, 
can explore the whole phase space \cite{Arnolddiff},\cite{Chirikov}. 
However, due to stickiness to higher dimensional invariant
tori Arnold diffusion can be suppressed so that 
certain stochastic regions are distinguished in which the trajectories
become trapped for longer times \cite{Kantz}. 

In fact, our findings imply that in the driven two-particle
system the motion takes place in ballistic channels
\cite{ballistic} associated with stickiness to two-dimensional invariant tori (cf. Fig.~\ref{fig:Fig3}). 
Notably, in the two
particle system the mean value $Q_x$ evolves in the same manner as
the single coordinate in the individual particle counterpart, viz. it exhibits effective growth. 
We performed numerical studies for  a large number of initial conditions contained in a potential well 
and found behavior equivalent to that 
illustrated in Figs.~\ref{fig:Fig1}-\ref{fig:Fig4}. 
The slight
alterations in the slope of the graph of $Q_x$ in Fig.~\ref{fig:Fig1} can
be explained with the fact that after a while the trajectory can
get released from a torus region to be captured in another one
where it again sticks to for a certain time. As noted above this release and capture 
behavior is manifested in the changes of the maximal amplitudes of
the oscillations of the variable $Q_y$ (see Figs.~\ref{fig:Fig1} and \ref{fig:Fig2}).

Alternatively to the geometrical phase space interpretation of the occurrence of directed motion supported by ballistic channels it is worthwhile to mention that from an analytical point of view the so called gating 
mechanism can be responsible for the initiation of the motion \cite{Mertens}. This mechanism is based on an interplay between an ac driving force and an ac parametric force causing dynamical symmetry breaking that may lead to directed motion. In fact Eq.~(\ref{eq:qx}) for the mean value variable $Q_x$ exhibits both features, i.e. it contains an ac driving term as well as the ac parametric force suggesting relevance of the gating mechanism.  

\vspace*{0.5cm}

\section{Extended Coupled Oscillator Systems}

Finally we consider extended coupled oscillator
chains where we aim to demonstrate that it is also possible to
generate directed motion with adiabatic periodic modulations of
the slope of the spatially periodic potential. 

Concerning the initial conditions we proceed as follows: Initially an amount
of energy $E_{n}=0.5
p_{n}^2+U_0(q_{n})<\Delta E$ is applied per unit such that the 
whole chain is elongated homogeneously along a fixed
position  $\tilde{q}_0$ near the bottom of the well. Then, the
position and/or momenta of all units are  
randomized.  The random position values are
chosen from a bounded interval $|q_n(0)-\tilde{q}_0| \leq \Delta q$ and,
likewise, the random initial momenta, $|p_n(0)-\tilde{p}_0|\leq \Delta p$. 
The whole chain is thus initialized close to an almost
homogeneous state, but yet sufficiently displaced  ($\Delta q \ne
0$) in order to generate nonvanishing interactions, entailing the
exchange of energy among the coupled units.

First we note that in the case without external
modulation of the slope there occurs the formation of a pattern of localized states due to modulational instability
(not shown here). Due the irregular dynamics it happens that
occasionally a unit overcomes the potential barrier but no
coordinated motion of the chain results. 

Remarkably, applying
the adiabatic modulation the entire chain not only escapes from the potential well but manages also
\begin{figure}
\includegraphics[scale=0.85]{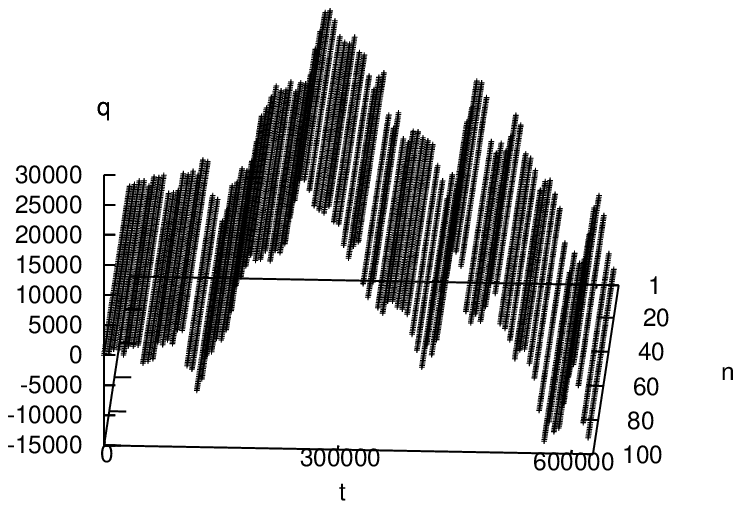}
\includegraphics[scale=0.85]{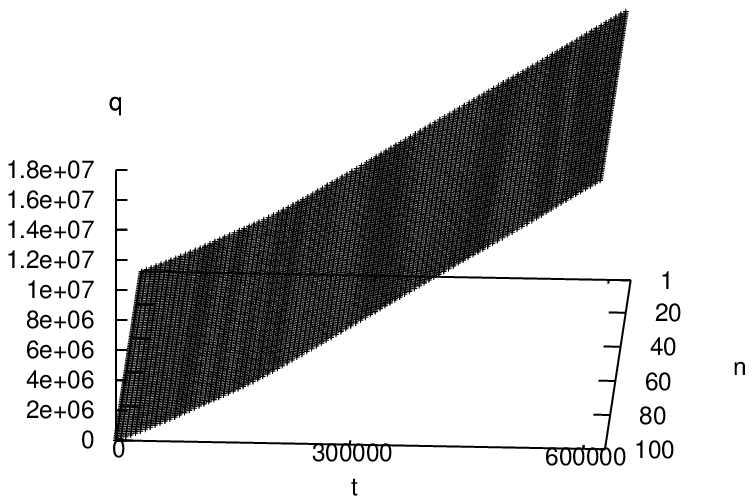}
\caption{Typical spatio-temporal evolution of the coordinates $q_n(t)$ for a chain of $100$ coupled particles. Initially the coordinates are uniformly distributed within the interval
$|q_n(0)-\tilde{q}_0|<\Delta q$ with an average $\tilde{q}_0=-0.35$ and  width $\Delta q=0.01$ and zero momenta, i.e. $p_n(0)=\tilde{p}_0=\Delta p=0$. The coupling strength is $\kappa=0.3$ and  the remaining parameter values are $F=0.05$, $\Omega=10^{-3}$ and $\Theta_0=0$.
Top (Bottom) panel: $\Omega=10^{-1}$ ($\Omega=10^{-3}$).}
\label{fig:Fig5}
\end{figure}
to travel freely and unidirectionally over a giant distance as seen from the spatio-temporal evolution 
of $q_n(t)$ in Fig.~\ref{fig:Fig5}.  In comparison for faster modulations no directed motion of the chain is obtained. The chain consists of $100$ coupled
oscillators and open boundary conditions are imposed.
The profile of the chain continuously undergoes changes with ensuing deviations from a flat state. 
Nevertheless the intriguing feature of transients of extremely long-range transport of the chain is provided by collective motion 
which is also reflected in the temporal behavior of the
\begin{figure}
\includegraphics[scale=0.45]{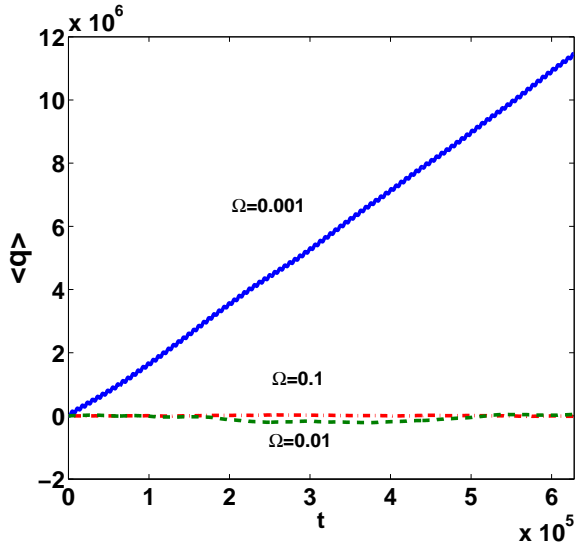}
\caption{Temporal evolution of the mean value $<{q}>$ for systems driven with different driving frequencies as indicated in the plot. Ensemble averages, denoted by $<\,\,\,>$ were performed over $100$ realizations of trajectories whose initial condition lie in the 
range given in Fig.~~\ref{fig:Fig5}. The parameter values are those given in Fig.~\ref{fig:Fig5}.}
\label{fig:Fig6}
\end{figure}
mean value of the coordinate $<{q}>=<\frac{1}{N}\sum_{n=1}^N\,q_n>$ shown in Fig.~\ref{fig:Fig6}. 
Ensemble averages, denoted by $<\,\,\,>$ were performed over $100$ realizations of trajectories whose initial condition lie in the 
range given in the caption of Fig.~\ref{fig:Fig5}.
Note that while for driving with 
$\Omega \lesssim 10^{-2}$ on average no directed motion results for sufficiently slow modulations (here illustrated for 
$\Omega=10^{-3}$) a large directed current is observed.
Conclusively, concerning transport the many particle system exhibits features equivalent to that 
observed in the system of individual particles \cite{Soskin},\cite{single}. 

It is interesting to relate our findings to those obtained for the continuum case of Eq.~(\ref{eq:qdot}), viz. a perturbed sine-Gordon equation that is driven by a mono-frequency periodic external field destroying integrability \cite{Olsen}. There also the emergence of a directed current for driven solitons results for suitable parameter values of the external field. In comparison we note that in the 1D continuum case a localized wave-like excitation in the form of a soliton moves  along the spatial coordinate whereas in our present study of the discrete counterpart the entire chain is coherently transported over the wells of the periodic potential landscape.

\section{Summary}

To summarize, we found that in the time-dependent driven Hamiltonian dynamics of a chain of linearly interacting particles evolving in
a symmetric, spatially periodic potential  directed transport can be induced. That is, when the slope of the potential is time-periodically varied
due to a slowly oscillating external modulation field the transport proceeds unidirectionally in a long-lasting transient period such that the chain covers huge distances.
The behavior of the mean value of the coordinates of extended chains is strikingly similar to that found in the system of noninteracting particles. 
Thus, the collective directed motion of the numerous microscopic degrees of freedom is manifested at a 
collective level in the evolution of the mean value of ${q}$.
We recall that for the system of individual particles 
transporting island structures corresponding to ballistic channels present in mixed phase space are responsible for transients of directed long-range transport. Our findings for the many particle system imply that ballistic channels exist in a high dimensional phase space too and support the collective motion of the coupled oscillators 
of a chain. For systems involving many degrees of freedom the issue of stickiness 
of invariant tori in high-dimensional phase space needs to be explored in more detail (an analysis which goes beyond the scope of the current article).

\end{document}